# Realize General Access Structure Based On Single Share

Yang Xie
School of Compute Science and Technology
University of Science and Technology of China
Hefei, China
xy080819@mail.ustc.edu.cn

Fuyou Miao
School of Compute Science and Technology
University of Science and Technology of China
Hefei, China
mfy@ustc.edu.cn

Sijjad Ali Khuhro
School of Compute Science and Technology
University of Science and Technology of China
Hefei, China
Sijjadali786@mail.ustc.edu.cn

Keju Meng
School of Compute Science and Technology
University of Science and Technology of China
Hefei, China
mkj@mail.ustc.edu.cn

***Abstract*—Traditional (*t,n*)-threshold secret sharing cannot realizing all access structures of secret sharing. So, Ito introduced the concept of Secret sharing scheme realizing general access structure. But Ito's scheme has to send multiple shares to each trustee. In this paper, we proposed two new secret sharing schemes realizing general access structures by only assigning one share to each trustee. Our proposed second scheme is a perfect secret sharing scheme. Furthermore, our schemes can realize any access structures.**

## I. Introduction

Secret sharing is an important topic in cryptography. The (*t,n*)-threshold secret sharing scheme (SSS) was first introduced by shamir [1] and Blakley [2] in 1979 separately. In (*t,n*)-threshold secret sharing, the secret is divided into n shares, each shares is distributed and retained by a participant. If t or more than t shares are gathered, the secret can be restored. And if less than t shares be gathered, no information can be obtained at all.

Secret sharing effectively reduces the risk of secret preservation and transmission; it has been widely used in many fields. Pedersen [3] proposed Distributed Key Generation (DKG) for key agreement based on secret sharing in 1991. Naor [4] proposed Visual Cryptography to improve the safety of visual sharing. SSS been used in digital signature, by distributed signature, the multiple subset of a group has the ability to generate the legal signature, and it also increases the difficulty of the attacker to forge the signature [5]. Secret sharing is also been used in Multi-Party Computation [6], Electronic Cash [7], key escrow [8], Electronic election [9] [10], Group Key [11] , and so on.

Although (*t,n*)-threshold SSS is almost perfect and widely used, sometimes, the access structure is not (*t,n*)-threshold. For example, there are 4 participants $p_1$, $p_2$, $p_3$, $p_4$ , a dealer wants to share a secret information with them. This dealer wants $p_1$ and $p_2$, $p_2$ and $p_3$, $p_1$ $p_3$ and $p_4$ can restore the secret. (2,4)-threshold or (3,4)-threshold cannot realize this requirement obviously. So SSS realizing general access structure (GAS) is proposed by Ito [12] in 1987.

Ito's scheme realized arbitrary access structure by assigning one or more shares to each participant. To reduce the number of distributed shares, Iwamoto [13], Li [14] and Harn [15] separately use (*t,n*)-threshold secret sharing realized general access structure by using integer programming or linear programming. Their scheme originates from weighted (*t,n*) secret sharing scheme proposed by Shamir [1]. In their scheme, each participant has it's positive weight. The secret can be recovered if the weight of participants equal or larger than the threshold.

However Iwamoto and Harn's scheme cannot realize all possible access structures, Li's scheme uses ramp assignment scheme [16][17] to reduce the shares assigned. Moreover, they also need to assign one or more shares to each participant.

Obviously, in these SSS for GAS, a most important issue is the number of shares distributed to each participant. However, in Ito's scheme, the number of distributed shares could be very large when the size of access structure is tremendous. On the other hand, Iwamoto and Harn's scheme also cannot reduce the number of distributed shares to an ideal size. What's more, there is not always has a solution for an integer programming problem.

More ever, many other schemes reduce the size of shares in special circumstances. Benaloh [18] proposed an efficient SSS for GAS if GAS can be described by a small monotone formula. And Karchmer[19] improve this scheme if this GAS can be described by a small monotone span program.

In this paper, we propose two new SSS realizing GAS by only assign one share to each trustee. Our proposed second scheme is perfect SSS. Furthermore, our schemes can realize any access structures.

The rest of this paper is organized as follows. In Section Ⅱ, we introduced some preliminaries, reference some classical definitions for SSS of GAS. In Section Ⅲ, introduce our first design, we implement SSS of GAS based on least common multiple (LCM). In Section Ⅳ, we proposed our second scheme based on minimal authorized subsets. In Section Ⅴ, we give analysis about our second scheme, and section Ⅵ conclude this paper.

## II. PRELIMINARIES

### A. Definitions

In a SSS, there are a set of trustees $P=\{p_1, p_2, \cdots, p_n\}$ and a dealer needs to share a secret among P. The Dealer decides which subset of trustees can reconstruct the secret together, we define the subset as authorized subset $A$. The dealer can also determine which subset of trustees cannot reconstruct the secret together, we define such a subset as unauthorized subset $B$. Traditionally, GAS never define sets with $A \subseteq B$, for example, if the dealer let $A = \{ p_1, p_2\}$ be an authorized subset, in GAS he cannot let $B = \{ p_1, p_2, p_3\}$ be a Unauthorized subset.

We define $A^-$ as minimal authorized subsets, and define $B^+$ as maximal unauthorized subsets. If there are two set of trustees $\Gamma$ and $\Gamma'$, all participants in $\Gamma'$ are also in $\Gamma$, and $\Gamma' \in A^-$, $\Gamma$ is an authorized subset and $\Gamma'$ is a member of minimal authorized subsets $A^-$. On the other hand, if there are two set of trustees $\Gamma$ and $\Gamma'$, all participants in $\Gamma$ are also in $\Gamma'$, and $\Gamma' \in B^+$, $\Gamma$ is a unauthorized subset and $\Gamma'$ is a member of maximal unauthorized subsets $B^+$.

So we concluded the following relationship：

$\Gamma$ is an authorized subset $A$ $\Leftrightarrow$ $\Gamma \supseteq \Gamma'$ and $\Gamma' \in A^-$

$\Gamma$ is a unauthorized subset $B$ $\Leftrightarrow$ $\Gamma \subseteq \Gamma'$ and $\Gamma' \in B^+$

### B. Shamir's (t,n)-threshold SSS

Shamir's SSS is widely used in secret sharing, it based on Lagrange interpolation polynomial. Shamir's SSS can be depicted as follows:

Shares Generation:

The Dealer chooses a large prime $p$, and a polynomial $f(x) = a_0 + a_1x + \cdots + a_{t-1}x^{t-1} \bmod p$ randomly, the secret $s = a_0 \in GF(p)$, and parameters $a_i \in GF(p)$, $(i = 1,\cdots,t-1)$. The dealer distributes the share $s_i = f(x_i)$ to $p_i$ $(i = 1,\cdots,n)$.

Secret Reconstruction:

If $m$ participants need to recover the secret s, they pool their shares together to compute the secret:

$$s = f(0) = \sum_{i=1}^{m} f(x_i) \prod_{j=1, j\neq i}^{m} \frac{-x_j}{x_i - x_j} (\bmod \ p)$$

If $m \geqslant t$, $m$ shares can recover the secret, else $m$ shares cannot get more information of the secret.

### C. Ito's SSS realizing GAS

In Ito's SSS for GAS. $P=\{p_1, p_2, \cdots, p_n\}$, let $|B^+|=k$, and $B^+=\{B_1, B_2, \cdots, B_k\}$. Construct $S=\{w_1, w_2, \cdots, w_k\}$ for $B^+$, $S$ is a set of shares constructed by $(k,k)$-threshold SSS. Then the secret can be restored when all $w_1, w_2, \cdots, w_k$ are obtained; and the secret cannot be restored if one of $w_i$ is missing.

For trustees $p_1, p_2, \cdots, p_n$, let the assignment formula be

$$g\left(p_j\right) = \{w_i : p_j \notin B_i \in B^+\} \ (j = 1,2,\ldots,n; i = 1,2,\ldots,k)$$

Example 1.

$$P = \{p_1, p_2, p_3, p_4\}$$
$$A^- = \{\{p_1, p_2, p_3\},\ \{p_2, p_3, p_4\},\ \{p_1, p_4\}\}$$

Then $B^+$ is given by

$$B^+ = \{\{p_1, p_2\},\ \{p_1, p_3\},\ \{p_2, p_3\}, \{p_2, p_4\}, \{p_3, p_4\}\}$$

For this $B^+$, we construct a (4,4)-threshold Shamir SSS, the shares are $S=\{w_1, w_2, w_3, w_4\}$. The assignments are given by

$$g\left(p_1\right) = \{w_3, w_4, w_5\}$$
$$g\left(p_2\right) = \{w_2, w_5\}$$
$$g\left(p_3\right) = \{w_1, w_4\}$$
$$g\left(p_4\right) = \{w_1, w_2, w_3\}$$

(End of Example.)

### D. Harn's SSS with GAS

Harn's scheme proposed a design to implement a GAS by integer programming. For participant set $P=\{p_1, p_2, \cdots, p_m\}$, let $A^-=\{A_1, A_2, \cdots, A_k\}$, $B^+=\{B_1, B_2, \cdots, B_l\}$. Define weight $W=\{w_1, w_2, \cdots, w_m\}$ is the number of shares assign to participants, $W(A_i)$ means the sum of participant's weight in set $A_i$ Then create integer programming inequations as following form:

$$n = min(w_1 + w_2 + \ldots + w_m)$$
$$W\left(A_i\right) > W\left(B_j\right) \ (i = 1,2,\ldots,k;\ j = 1,2,\ldots,l)$$
$$w_i \geq 1 \ (i = 1,2,\ldots,m)$$

If these inequations have solution, $w_i$ can be determined. The threshold

$$t = min\{W\left(A_1\right), W\left(A_2\right), \ldots, W\left(A_k\right)\}$$

Then, the dealer construct $(t,n)$-threshold SSS by any classical SSS. Assign shares as the number of $w_i$ to $p_i$. Then the secret can be restored when more than $t$ shares are obtained; and the secret cannot be restored from less than $t$ shares. $W(A_i) \geqslant \min\{W(A_i)\} = t > W(B_j)$ $(i=1,2,\ldots,k; j=1,2,\ldots,l)$. So SSS of GAS can be realized.

Example 2.

$$P = \left\{p_1, p_2, p_3, p_4\right\}$$

$$A^- = \left\{\left\{p_1, p_2\right\}, \left\{p_2, p_3\right\}, \left\{p_1, p_3, p_4\right\}\right\}$$

Then $B^+$ is given by

$$B^+ = \{\{p_1, p_3\}, \{p_1, p_4\}, \{p_2, p_4\}, \{p_3, p_4\}\}$$

From this $A^-$ and $B^+$, we can establish the following inequality conditions:

$$w_1 + w_2 > \{w_1 + w_3, w_1 + w_4, w_2 + w_4, w_3 + w_4\}$$
$$\Rightarrow w_2 > w_3, w_2 > w_4, w_1 > w_4$$
$$w_2 + w_3 > \{w_1 + w_3, w_1 + w_4, w_2 + w_4, w_3 + w_4\}$$
$$\Rightarrow w_2 > w_1, w_2 + w_3 > w_1 + w_4, w_3 > w_4, w_2 > w_4$$
$$w_1 + w_3 + w_4 > \{w_1 + w_3, w_1 + w_4, w_2 + w_4, w_3 + w_4\}$$
$$\Rightarrow w_1 + w_3 > w_2$$

In summary, we have:

$$w_2 > w_3 > w_4, w_2 > w_1 > w_4$$
$$w_2 + w_3 > w_1 + w_4, w_1 + w_3 > w_2$$

So we can obtain $w_1$=3, $w_2$=4, $w_3$=2, $w_4$=1 and $t$ = 6.

Construct (6,10)-threshold SSS, and assign shares by weight.

(End of Example.)

But sometimes, integer programming may not have solution. Harn's scheme cannot realize these GAS. For example, $P = \{p_1, p_2, p_3, p_4\}$, $A^- = \{\{p_1, p_2\}, \{p_3, p_4\}, \{p_1, p_3\}\}$.

## III. REALIZE SSS OF GAS BY LCM

For a GAS, all the schemes mentioned above assign multiple shares to participants. In SSS of GAS, an important issue is to reduce the number of shares distributed to each participant. So we propose a scheme only distribute one share to each trustee.

A dealer wants to share secret $S$ among trustees $P=\{p_1, p_2, \cdots, p_n\}$, let $A^-=\{A_1, A_2, \cdots, A_l\}$, $B^+=\{B_1, B_2, \cdots B_k\}$, the scheme can be described as follows.

Shares Generation:

1. The dealer selects a prime number $q$, and choose a sequence of pairwise coprime positive integers $q_1, q_2, \cdots, q_k$, to let

$$q_1 * q_2 * \ldots * q_k \bmod q = S, 0 < q_i < q \; (i = 1, 2, \ldots, k)$$

2. The dealer computes the share $sh_j$ for each trustee.

$$sh_j = \prod_{p_j \notin B_i} q_i \; (j = 1, 2, \ldots, n; \; i = 1, 2, \ldots, k)$$

3. Assign $sh_j$ to each trustee $p_j$, broadcast $q$.

Secret Reconstruction:

If some trustees (they constitute a set $\Gamma$) need to recover the secret $S$, they use their shares $sh_i$ to compute the secret as

$$S = LCM\{sh_i : p_i \in \Gamma\} \bmod q$$

Example 3.

$$P = \{p_1, p_2, p_3, p_4\}$$

$$A^- = \{\{p_1, p_2\}, \{p_2, p_3\}, \{p_1, p_3, p_4\}\}$$

Then $B^+$ is given by

$$B^+ = \{\{p_1, p_3\}, \{p_1, p_4\}, \{p_2, p_4\}, \{p_3, p_4\}\}$$

The Dealer selects and broadcasts $q$ to all trustees, chooses a sequence of pairwise coprime positive integers $q_1, q_2, q_3, q_4$, to let $q_1 * q_2 * q_3 * q_4 \bmod q = S$

Then compute the $sh_i$ for each $p_i$:

$$sh_1 = q_3 * q_4$$
$$sh_2 = q_1 * q_2 * q_4$$
$$sh_3 = q_2 * q_3$$
$$sh_4 = q_1$$

When $\Gamma = \{p_1, p_2, p_3\}$ want to recover the secret, LCM$\{sh_1, sh_2, sh_3\} = q_1 * q_2 * q_3 * q_4$. They can reconstruct the secret.

When $\Gamma = \{p_1, p_3\}$ want to recover the secret, LCM$\{sh_1, sh_3\} = q_2 * q_3 * q_4$. They cannot reconstruct the secret.

(End of Example.)

By using LCM to combine multiple $q_i$ to a single share, we realized a scheme where only a single share is assigned to realize GAS. But there are two other problems.

Firstly, the worst information rate [20] (the ratio between the maximum size of the share and the size of secret) of it is obviously bad. The size of share combined by $q_i$ has the same size with these multiple $q_i$.

Secondly, it is a ramp SSS. Suppose $S = q_1 * q_2 * \cdots * q_k \bmod q$, when the participants of $B$ got $q_1, q_2, \cdots, q_{k-1}$, they can know $q_k$ can only be the number that makes GCD($q_k$, $q_i$)=1 (i=1,2,$\cdots$,$k$-1; GCD means Greatest Common Divisor). So the secret they could be supposed is no longer in the range of 0 ~ $q$-1.

In conclusion, we should reduce the size of share into the same field of secret, and remove the relevance of shares.

But how to solve these problems?

If all $q_i$ in one share can be inferred by one of $q_i$, the size of share can be reduced to $q$. And if the $q_i$ needs not to be coprime positive integers, the range of secret cannot be reduced by the known shares neither.

So we propose the second scheme realizing SSS of GAS by broadcasting information.

## IV. REALIZING SSS OF GAS BY BROADCASTING INFORMATION

This scheme can be described as follows.

A dealer wants to share secret $S$ for a trustees set $P=\{p_1, p_2, \cdots, p_n\}$, let $A^- = \{A_1, A_2, \cdots, A_k\}$, we need not to know $B^+$.

**Shares Generation:**

1. The dealer selects a integer $m$ that $S<2^m$, and randomly choose n integers $q_1, q_2, \cdots, q_n$ from 0 to $2^m$.
2. For each authorized subset $A_i$ ($i$=1,2, $\cdots$ ,$k$), dealer construct

$$w_{i1} \oplus w_{i2} \cdots \oplus w_{in} = S \ (\oplus is\ XOR\ symbol)$$

if $p_j \in A_i$ . $0 < w_{ij} < 2^m$, else $w_{ij} = 0$.

3. Construct a matrix $D$ by $w_{ij}$ and $q_j$, the element $d_{ij}$ in the $i$ th row and $j$ th column of $D$ is evaluated as follows.

$$d_{ij} = \begin{cases} w_{ij} + q_i \left( mod\ 2^m \right) & w_{ij} \neq 0 \\ 0 & w_{ij} = 0 \end{cases}$$

4. Assign $q_i$ to each trustee $p_i$, broadcast $D$ and $m$.

**Secret Reconstruction:**

1. If a trustees set $\Gamma$ need to recover the secret $S$, the trustees $p_i \in \Gamma$ use their shares $p_i$ and the public matrix $D$ to compute their $w_{ij}$ .

$$w_{ij} = \begin{cases} d_{ij} + 2^m - q_i \left( mod\ 2^m \right) & d_{ij} \neq 0 \\ 0 & d_{ij} = 0 \end{cases}$$

2. If $\Gamma \supseteq A_i$ for $A_i \in A^-$, trustees in $\Gamma$ can get enough $w_{ij}$ to recover the secret $S$, or they cannot get any more information of $S$.

Example 4.

$$P = \{p_1, p_2, p_3, p_4\}$$

$$A^- = \{\{p_1, p_2\},\ \{p_1, p_3\},\ \{p_2, p_3, p_4\}\}$$

Dealer randomly choose and send $q_i$ to each $p_i$ ($i$=1,2,3,4). Choose integer $w_{11}$, $w_{12}$, $w_{21}$, $w_{23}$, $w_{32}$, $w_{33}$, $w_{34}$ let

$$w_{11} \oplus w_{12} = S$$

$$w_{21} \oplus w_{23} = S$$

$$w_{32} \oplus w_{33} \oplus w_{34} = S$$

Then compute the $D$ matrix.

$$D = \begin{bmatrix} w_{11} + q_1 (\bmod 2^m) & w_{12} + q_2 (\bmod 2^m) & 0 & 0 \\ w_{21} + q_1 (\bmod 2^m) & 0 & w_{23} + q_3 (\bmod 2^m) & 0 \\ 0 & w_{32} + q_2 (\bmod 2^m) & w_{33} + q_3 (\bmod 2^m) & w_{34} + q_4 (\bmod 2^m) \end{bmatrix}$$

Broadcast the matrix $D$, so the participants can know which subset can be used to recover the secret.

If $\Gamma = \{p_1, p_2\}$ want recover $S$, they compute their share

$$w_{11} = d_{11} + 2^m - q_1 \left( mod\ 2^m \right)$$

$$w_{12} = d_{12} + 2^m - q_2 \left( mod\ 2^m \right)$$

Then they can get the secret from $w_{11} \oplus w_{12} = S$.

If $\Gamma = \{p_2, p_3\}$ want to recover $S$, they cannot find an authorized subset row to compute their shares, so they can only know $0 < S < 2^m$.

(End of Example.)

## V. ANALYSES AND DISCUSSIONS FOR THE SECOND SCHEME

### A. *Security Analyses*

In our proposed second scheme, secret $S$ is constructed by several $w_{ij}$ XOR, $S$ cannot be reconstructed if one of them missing.

The range of $S$ is in 0 ~ $2^m$-1, and because $w_{ij}$ is in the range of 0 ~ $2^m$-1, so that a set of unauthorized subset cannot get more information about the secret $S$. They can only suppose the secret from 0 ~ $2^m$-1. So we can assert the proposed second scheme is a perfect secret sharing scheme.

The public matrix $D$ is also unsolvable without $q_i$.

We use an example to demonstrate it.

Example 5.

$$P = \{p_1, p_2, p_3\}$$

$$A^- = \{\{p_1, p_2\},\ \{p_1, p_3\},\ \{p_2, p_3\}\}$$

We can use SSS of GAS to realize ($t,n$)- threshold SSS.

Dealer randomly choose and send $q_i$ to each $p_i$ ($i$=1,2,3). Choose integer $w_{11}$, $w_{12}$, $w_{21}$, $w_{23}$, $w_{32}$, $w_{33}$ let

$$w_{11} \oplus w_{12} = S$$

$$w_{21} \oplus w_{23} = S$$

$$w_{32} \oplus w_{33} = S$$

Then compute the $D$ matrix.

$$D = \begin{bmatrix} w_{11} + q_1 (\bmod 2^m) & w_{12} + q_2 (\bmod 2^m) & 0 \\ w_{21} + q_1 (\bmod 2^m) & 0 & w_{23} + q_3 (\bmod 2^m) \\ 0 & w_{32} + q_2 (\bmod 2^m) & w_{33} + q_3 (\bmod 2^m) \end{bmatrix} = \begin{bmatrix} d_{11} & d_{12} & 0 \\ d_{21} & 0 & d_{23} \\ 0 & d_{32} & d_{33} \end{bmatrix}$$

Assume $p_2$ want recover the $S$ from $q_2$ and $D$.

$p_2$ can get $w_{12}$ and $w_{32}$ from $D$, then he can get

$$\left( d_{11} - q_1 \right) \oplus w_{12} = S = \left( d_{33} - q_3 \right) \oplus w_{32}$$

$$\Rightarrow \left( d_{11} - q_1 \right) \oplus \left( d_{33} - q_3 \right) = w_{12} \oplus w_{32}$$

And from the second row of $D$, he can get

$$\left( d_{21} - q_1 \right) \oplus \left( d_{23} - q_3 \right) = S$$

Even though $d_{11}$, $d_{33}$, $w_{12}$, $w_{32}$, $d_{21}$, $d_{23}$ are already known. $p_2$ still cannot compute $q_1$ and $q_3$.

Because $(a + b) \oplus c = a \oplus c + b \oplus c$, only if $a+b=a \oplus b$.

XOR is not in the same field of addition. So $p_2$ can only assume $q_1$ or $q_3$ from 0 ~ $2^m$-1, thus he can only assume $S$ from 0~$2^m$-1.

### B. *Efficiency analyses*

The previous schemes are mostly based on maximal unauthorized subsets, but the maximal unauthorized subsets is not always been given. It also needs computation from minimal authorized subsets. Our proposed second scheme construct shares and public matrix from minimal authorized subsets, it's obviously more efficient than the previous schemes under these circumstances.

In our proposed second scheme, the size of secret is $2^m$, and the size of share is $2^m$ too. So the information rates of the second scheme is $\rho$=1.

In the previous schemes realizing GAS by Shamir's SSS, the most time-consuming operation in them is the polynomial interpolation computation. But in this paper, the second scheme realizing GAS only by simple addition, XOR and modular operation. Therefore, this scheme is very efficient and easy to implement.

## VI. CONCLUSION

This paper has proposed two new SSS of GAS. The first scheme realizes GAS by only assigning single share based on LCM, but there are some problem with it. So we proposed the second one, the second scheme is perfect SSS and only assign one share to each participant. Furthermore, for any authorized subsets, the second scheme is more efficient than previous schemes.

## ACKNOWLEDGMENT

This research work was supported by the National Natural Science Foundation of China under 61572454, 61572453, 61520106007.